\documentclass[12pt]{article}
\usepackage{amssymb}
\usepackage{amsmath}
\usepackage{graphicx}
\begin{document}
\title{Tachyon inflation in an AdS braneworld with back-reaction}

\author{
Neven Bili\'c$^1$\thanks{bilic@irb.hr}, Dragoljub Dimitrijevic$^2$\thanks{ddrag@pmf.ni.ac.rs}, 
Goran Djordjevic$^2$\thanks{gorandj@junis.ni.ac.rs},
 and  Milan Milosevi\'c$^2$\thanks{mmilan@seenet-mtp.info},
  \\
$^1$Division of Theoretical Physics, Rudjer Bo\v{s}kovi\'{c} Institute, Zagreb, Croatia\\
$^2$Department of Physics,
University of Ni\v s,  Srbija\\
}

\maketitle

\begin{abstract}
We analyze the inflationary scenario  based on the tachyon 
field coupled with the radion of the second Randall-Sundrum model (RSII).
The tachyon Lagrangian is derived from the dynamics of a 3-brane moving in the 
five dimensional bulk.
The AdS$_5$  geometry of the bulk is extended to include the radion.
Using the Hamiltonian formalism we find four nonlinear field equations
supplemented by the modified Friedmann equations of
the RSII braneworld cosmology.
After a suitable rescaling we reduce the parameters of our model to only one free parameter
related to the brane tension and the AdS$_5$ curvature.
We solve the equations numerically  assuming a reasonably wide range of initial conditions
determined by physical considerations.
Varying the free parameter  and  initial conditions
we confront our results with
the  Planck 2015 data.
\end{abstract}


\section{Introduction}	
The inflationary universe scenario \cite{starobinsky,guth,linde} in which the early universe undergoes a rapid expansion
has been generally accepted as a solution to the horizon problem and some other 
related problems of the standard big-bang cosmology.
The origin of the field that drives inflation is still unknown and is subject to speculations.
Among many models of inflation a popular class comprise tachyon inflation models
\cite{fairbairn,feinstein,shiu1,kofman,sami,shiu2,cline,steer,campo,li,tachyon}.
These models are of particular interest as in these models inflation 
is driven by the tachyon field originating in string theory.

The tachyon potential is
derived from string theory and has to satisfy some definite properties to describe
tachyon condensation and other requirements in string theory.  
However, Kofman and Linde have  shown \cite{kofman} that the slow-roll conditions are not
compatible with a string coupling much smaller than one, and the compactification length scale much 
larger than the Planck length.
This leads to the density fluctuations produced during inflation being incompatible
with observational constraint on the amplitude of the scalar perturbations.
This criticism is based on the string
theory motivated values of the parameters in the tachyon potential, i.e., the brane tension
and the parameters in the four-dimensional Newton constant obtained
via conventional string compactification. Of course, if one relaxes 
the string theory constraints on the above mentioned parameters,
the effective tachyon  theory will naturally lead to a type of inflation
which will slightly deviate from the conventional inflation based on
the canonical scalar field theory.
Steer and Vernizzi \cite{steer} have noted a deviation from  the standard single field inflation
in the second order consistency relations. Based on their analysis they concluded that
the tachyon inflation could not be ruled out by the then available observations.
It seems like the present observations \cite{planck2015}
 could perhaps discriminate between different tachyon models
and disfavor or rule out some of these models
(for a recent discussion on phenomenological constraints
imposed by Planck 2015, see, e.g., ref \cite{pirtskhalava}).

A simple tachyon model can be analyzed in the framework of the second Randall-Sundrum (RSII)
model \cite{randall2}. The RSII model was originally proposed as a possible mechanism for localizing gravity
on the 3+1 universe embedded in a 4+1 dimensional spacetime without compactification of the extra dimension.
The model is a 4+1 dimensional Anti de Sitter (AdS$_5$) universe containing two 3-branes with opposite tensions
separated in the fifth dimension: observers reside on the positive tension brane 
and the negative tension brane is pushed off to infinity.
The Planck mass scale is determined by the curvature of the AdS spacetime rather then 
by the size of the fifth dimension.

The fluctuation of the interbrane distance  along the extra dimension implies the existence of 
the so called {\em radion} -- a massless scalar field that causes a distortion of the bulk geometry.
In this regard, a stabilization mechanism of the interbrane distance has been proposed \cite{goldberger}
by assuming the presence of scalar fields in the bulk.
The stabilization mechanism is relevant for the RSI model where the interbrane distance is kept
finite. In RSII model, as the negative tension brane is pushed off to infinity
the radion disappears.  However, it has been shown by Kim, Tupper, and Viollier \cite{kim}
that a disappearance of the radion in RSII is an artifact of linear theory
and hence, when going beyond linear theory  the radion remains a dynamical field
in the RSII model.
Moreover, owing to the radion,  
the distance between branes remains finite in the RSII limit of infinite 
coordinate bulk even though the coordinate position of the second brane is infinite.

The presence of the radion may have interesting physical implications.
The radion field has been proposed as an inflaton \cite{barvinsky}.
It has been shown \cite{bilic2013b,bilic2013a} that the interaction of the radion with the tachyon can alter the 
tachyon equation of state: by averaging over large scales the effective equation of state
describes the warm dark matter. One of our aims here is to answer the question
whether radion could drastically change the early cosmology in the framework of the RSII 
braneworld model.

In this paper we propose a simple tachyon condensate as a model for inflation
and  we analyze two effects:
the coupling of the tachyon with the radion  and the modification of the
standard  cosmology in
the RSII scenario. 
To study these effects we will consider an additional 
dynamical 3-brane moving in the AdS$_5$ background  of the RSII model. 
The action of the 3+1 dimensional brane in the five dimensional bulk is equivalent 
to the Dirac-Born-Infeld (DBI) description of the Nambu-Goto 3-brane.
\cite{bordemann,jackiw}.
It is a simple matter to show that this additional 3-brane 
behaves effectively as a tachyon with the inverse quartic potential \cite{bilic2013a}.
The tachyon model of this kind falls into the class of the power-law tachyon potentials 
$V(\theta) \propto \theta^{-n}$ with $n>2$ which drive a dark-matter  attractor \cite{abramo}:
for $\theta \rightarrow \infty$, the pressure tends to unity very quickly with the unpleasant feature 
of cold dark matter (CDM) domination at the end of inflation.
Actually, this problem is imminent for all tachyon models with the ground state at 
$\theta\rightarrow\infty$ \cite{kofman}. The tachyon field rolls towards its ground state  
without oscillating about it and the conventional reheating mechanism does not work.
Typically these scenarios
suffer from a reheating problem, since gravitational
particle production is not efficient compared to
the standard non-gravitational particle production by an
oscillating inflaton field. This will be discussed in more detail at the end of Sec.\ \ref{conditions}.

The remainder of the paper is organized as follows. 
In Sec. \ref{Sec:Randal} we present 
the RSII model with back-reaction. The system of dynamical equations 
is presented in a dimensionless form suitable for the calculation of the expansion rate and slow-roll parameters. 
In Sec. \ref{conditions}
the slow-roll approximation and initial conditions for our model are discussed in detail.
Numerical results are presented and discussed in Sec. \ref{Sec:numres}.
In the concluding section, Sec.\ \ref{conclusions}, we summarize our results and give conclusions.

\section{Randall-Sundrum model with back-reaction}\label{Sec:Randal}
The bulk spacetime of the extended RSII model  which includes
the back-reaction of the radion  \cite{kim} in Fefferman-Graham coordinates is described by the 
line element
\begin{equation}
ds^2_{(5)}=G_{ab} dX^a dX^b=\frac{1}{k^2 z^2}\left[\left(1+ k^2 z^2\eta(x)\right)g^{\mu\nu}dx^\mu  dx^\nu
  -\frac{1}{\left(1+ k^2 z^2\eta(x)\right)^2} dz^2\right] ,
 \label{eq2001}
\end{equation}
where $k=1/\ell$ is the inverse of the
AdS curvature radius $\ell$ and 
$\eta (x)$ is the radion field.
The observer brane is placed at $z=\ell$ and $g_{\mu\nu}$ is the metric on the brane.

Consider a dynamical 3-brane moving in the bulk. 
The brane Lagrangian describes 
the dynamic of the tachyon field modified by the interaction with the radion.
After integrating out the fifth coordinate the total effective  action is given by \cite{kim}
\begin{equation}
S = \int d^4x \sqrt{-g}
\left( -\frac{R}{16\pi G}+\frac12
g^{\mu\nu}\Phi_{,\mu}\Phi_{,\nu} 
\right)+ S_{\rm br} ,
\label{eq3015}
\end{equation}
where $\Phi$ is the canonically normalized radion field related to $\eta$ as
\begin{equation}
\eta = \sinh^2 \left( \sqrt{\frac{4\pi G}{3}} \Phi \right).
\end{equation}
The brane action $S_{\rm br}$  is derived in Ref.\ \cite{bilic2013a}
in terms of 
the induced  metric or  the ``pull back" of the bulk space-time metric 
$G_{ab}$ to the brane,
\begin{equation}
g^{(\rm ind)}_{\mu\nu}=G_{ab}
\frac{\partial X^a}{\partial x^\mu}
\frac{\partial X^b}{\partial x^\nu} \, .
\label{eq0002}
\end{equation}
The action takes the form
\begin{equation}
S_{\rm br} = -\sigma  \int d^4x \sqrt{- \det g^{\rm ind}_{\mu \nu}} 
 =- \int  d^{4}x  \sqrt{-g} \,
\frac{\sigma}{k^4\Theta^4}(1+k^2\Theta^2 \eta)^2
\sqrt{1-\frac{g^{\mu\nu}\Theta_{,\mu}\Theta_{,\nu}}{(1+k^2\Theta^2\eta)^3}},
 \label{eq2006}
\end{equation}
where $\sigma$ denotes the brane tension 
and $\Theta$ is the tachyon field.

In the absence of the radion ($\Phi = 0$)  
the brane action is just the tachyon condensate with the inverse quartic potential
\begin{equation}
S_{\rm br}^{(0)} =
 - \int  d^{4}x  \sqrt{-g} \frac{\lambda}{\Theta^4} \sqrt{1-g^{\mu\nu}\Theta_{,\mu}\Theta_{,\nu}},
\label{tachS}
\end{equation}
where
\begin{equation}
\lambda = \frac{\sigma}{k^4}.
\end{equation}

The combined brane-radion Lagrangian reads
\begin{equation}
{\cal{L}} = \frac12 g^{\mu\nu}\Phi_{,\mu}\Phi_{,\nu}  
-\frac{\lambda\psi^2}{\Theta^4}\sqrt{1-\frac{g^{\mu\nu}\Theta_{,\mu}\Theta_{,\nu}}{\psi^3}},
 \label{lagrangian}
\end{equation}
where
\begin{equation}
\psi=1+k^2 \Theta^2 \eta .
\end{equation}
Hence, in our model, the tachyon is as usual the Dirac-Born-Infeld type scalar field whereas 
the radion is a canonical scalar.
In the following we will assume the spatially flat FRW spacetime on the observer brane
with four dimensional line element in the standard form
\begin{equation}
 ds^2=g_{\mu\nu}dx^\mu dx^\nu=dt^2-a^2(t)(dr^2+r^2 d\Omega^2) .
 \label{eq0012}
\end{equation}

The treatment of our system in a cosmological context is conveniently performed in the Hamiltonian formalism.
For this purpose we first define 
the conjugate momentum fields as
\begin{equation}
\Pi_\Phi^\mu=
\frac{\partial{\cal{L}}}{\partial\Phi_{,\mu}},
\label{eq2115}
\quad
\Pi_\Theta^\mu=
\frac{\partial{\cal{L}}}{\partial\Theta_{,\mu}}.
\end{equation}
In the cosmological context    $\Pi_\Phi^\mu$    and  $\Pi_\theta^\mu$ are time-like so we may also define 
their magnitudes as 
\begin{equation}
\Pi_\Phi=\sqrt{g_{\mu\nu}\Pi_\Phi^\mu\Pi_\Phi^\nu} , 
\hspace{1cm}
\Pi_\theta=\sqrt{g_{\mu\nu}\Pi_\theta^\mu\Pi_\theta^\nu}.
\label{eq2118}
\end{equation}
The Hamiltonian density may  be derived from the stress tensor corresponding to the
Lagrangian (\ref{lagrangian}) or by the Legendre transformation.
Either way one finds \cite{bilic2013a}
\begin{equation}
{\cal{H}} = \frac{1}{2}\Pi_{\Phi}^2+\frac{\lambda\psi^2}{\Theta^4}\sqrt{1+\Pi_{\Theta}^2\Theta^8/(\lambda^2\psi)}.
 \label{totH}
\end{equation}
For later use we may also need the Hamiltonian density in terms of  $\Phi_{,\mu}$ and $\Theta_{,\mu}$,
\begin{equation}
{\cal{H}} = \frac12 g^{\mu\nu}\Phi_{,\mu}\Phi_{,\nu}  
+\frac{\lambda\psi^2}{\Theta^4}\left(1-\frac{g^{\mu\nu}\Theta_{,\mu}\Theta_{,\nu}}{\psi^3}\right)^{-1/2}.
 \label{eq014}
\end{equation}

Next, we can write Hamilton's equations in the form
\begin{eqnarray}
\dot{\Phi} = \frac{\partial{\cal{H}}}{\partial\Pi_\Phi},\label{eqHam1}\\
\dot{\Theta} = \frac{\partial{\cal{H}}}{\partial\Pi_\Theta},\label{eqHam2} \\
\dot{\Pi}_\Phi +3H\Pi_\Phi=-\frac{\partial{\cal{H}}}{\partial\Phi},\label{eqHam3}\\
\dot{\Pi}_\Theta + 3H\Pi_\Theta=-\frac{\partial{\cal{H}}}{\partial\Theta}.
\label{eqHam4}
\end{eqnarray}
In the spatially flat Randall-Sundrum cosmology 
the Hubble expansion rate $H$ is related to the Hamiltonian via the
modified Friedmann equation \cite{maartens}
\begin{eqnarray} 
 H\equiv\frac{\dot{a}}{a}=\sqrt{\frac{8 \pi G}{3} \mathcal{H}\left(1+ \frac{2 \pi G}{3k^2} \mathcal{H}\right) }.
 \label{scale_a}
\end{eqnarray}
In addition, we will make use of the energy-momentum  conservation equation
\begin{equation}
\dot{\mathcal{H}}+3H(\mathcal{H}+\mathcal{L})=0 ,
 \label{3200}
\end{equation}
which, combined with the time derivative of (\ref{scale_a}),
yields the second Friedmann equation in the form
\begin{equation}
\dot{H}=-4\pi G(\mathcal{H}+\mathcal{L})\left(1+
\frac{4 \pi G}{3k^2} \mathcal{H}\right).
 \label{eq3222}
\end{equation}
Thus, the Friedman equations are modified in the RSII cosmology.
As far as we know the effects of these modifications on tachyon inflation were first studied 
by Bento, Bertolami and Sen \cite{bento}.

To solve the system of equations (\ref{eqHam1})-(\ref{scale_a}), it is convenient to rescale the time as $t=\tau/k$ and  express  
the system in terms of dimensionless quantities. 
Besides, we can eliminate  the coupling constant $\lambda$ from the equations by appropriately rescaling
the fields $\Phi$ and $\Theta$ and their  conjugate fields $\Pi_{\Phi}$ and
 $\Pi_{\Theta}$.
To this end  we  introduce  the dimensionless functions 
\begin{eqnarray}
h = H/k, \quad 
\phi = \Phi/(k \sqrt{\lambda}),\quad 
\pi_\phi =  \Pi_\Phi/(k^2\sqrt{\lambda})),\quad
\theta=k \Theta, \quad
\pi_\theta = \Pi_{\Theta}/(k^4 \lambda),
\label{eq002}
\end{eqnarray}
and rescale the Lagrangian and Hamiltonian to obtain the 
dimensionless
pressure and energy density:
\begin{equation}
\bar{p}\equiv \frac{\mathcal{L}}{k^4\lambda} =\frac{1}{2}\pi_\phi^2-\frac{\psi^2}{\theta^4} 
\frac{1}{\sqrt{1+\theta^8\pi_\theta^2/\psi}} ,
\label{eq504}
\end{equation}
\begin{equation}
\bar{\rho}\equiv \frac{\mathcal{H}}{k^4\lambda}=\frac{1}{2}\pi_\phi^2+\frac{\psi^2}{\theta^4} 
\sqrt{1+\theta^8\pi_\theta^2/\psi}. 
\label{eq505}
\end{equation}
Following Steer and Vernizzi \cite{steer} we also introduce a combined dimensionless coupling
\begin{equation}
\kappa^2=8\pi\lambda G k^2 .
\label{eq102}
\end{equation}
Then, from (\ref{eqHam1})-(\ref{scale_a})
we obtain the following set of equations 
\begin{equation}
\label{sysRT1}
\dot \phi=\pi_\phi,
\end{equation}
\begin{equation}
\label{sysRT2}
\dot \theta=\frac{\theta^4\psi\pi_{\theta}}
{\sqrt{1+\theta^8\pi_{\theta}^2/\psi}},
\end{equation}
\begin{equation}
\label{sysRT3}
\dot \pi_\phi=-3h\pi_\phi
-\frac{\psi}{2\theta^2}\,\frac{4+3\theta^8\pi_{\theta}^2/\psi} 
{\sqrt{1+\theta^8\pi_{\theta}^2/\psi}}\eta',
\end{equation}
\begin{equation}
\label{sysRT4}
\dot \pi_{\theta}=-3h{{\pi }_{\theta}}
+\frac{\psi}{\theta^5}\,\frac{4-3\theta^{10}\eta\pi_{\theta}^2/\psi} 
{\sqrt{1+\theta^8\pi_{\theta}^2/\psi}},
\end{equation}
where
\begin{eqnarray}
\label{h}
h\equiv \frac{\dot{a}}{a}=\sqrt{\frac{\kappa^2}{3}\bar{\rho}\left(1+\frac{\kappa^2}{12}\bar{\rho}  \right)},
\end{eqnarray}
\begin{equation}
\psi=1+\theta^2 \eta,
\end{equation}
\begin{equation}
\eta = \sinh^2 \left( \sqrt{\frac{\kappa^2}{6}} \phi \right) ,
\end{equation}
\begin{equation}
\eta^{\prime}=\frac{d\eta}{d\phi} 
= \sqrt{\frac{\kappa^2}{6}} \sinh \left( \sqrt{\frac{2 \kappa^2}{3}} \phi \right).
\label{eq012}
\end{equation}
In addition to Eqs.\ (\ref{sysRT1})-(\ref{h}) we can solve in parallel 
the second Friedman equation 
\begin{eqnarray}
\dot{h}=-\frac{\kappa^2}{2}(\bar{\rho}+\bar{p})\left(1+\frac{\kappa^2}{6}\bar{\rho}  \right),
\label{h2}
\end{eqnarray}
and 
\begin{equation}
\dot{N}=h
 \label{n}
\end{equation}
where $N$ is the number of e-folds as a function of $\tau$.

In Eqs.\ (\ref{sysRT1})-(\ref{n}) and from now on the overdot denotes a derivative with respect to $\tau$.
Obviously, the explicit dependence on $\lambda$ in Eqs.\ (\ref{sysRT1})-(\ref{eq012}) is eliminated and 
 the only remaining free parameter is $\kappa$.
Equations (\ref{sysRT1}) and (\ref{sysRT2}) can be used to express the pressure and energy density
in terms of $\dot{\phi}$ and  $\dot{\theta}$. One thus finds
\begin{equation}
\bar{p}=\frac{1}{2}\dot{\phi}^2-\frac{\psi^2}{\theta^4} 
\sqrt{1-\dot{\theta}^2/\psi^3} ,
\label{eq015}
\end{equation}
\begin{equation}
\bar{\rho}=\frac{1}{2}\dot{\phi}^2+\frac{\psi^2}{\theta^4} 
\frac{1}{\sqrt{1-\dot{\theta}^2/\psi^3}}.
\label{eq016}
\end{equation}


The functional dependence of $h$ on $\tau$ will be used
to calculate the  slow-roll parameters
and the number of e-folds.
The slow-roll parameters are defined as \cite{steer,schwarz} 
\begin{equation}
\epsilon_{i} \equiv \frac{d\ln| \epsilon_{i-1}|}{Hdt},\qquad i \geq 1,
\end{equation}
where
\begin{equation}
\epsilon_0 \equiv \frac{H_*}{H}.
\end{equation}
and $H_*$ is the Hubble rate at an arbitrarily chosen time. 
Using the previously defined dimensionless Hubble rate $h$, the first two parameters can be written as
\begin{equation}\label{eps1}
\epsilon_1 = -\frac{\dot{h}}{h^2},
\end{equation}
\begin{equation}\label{eps2}
\epsilon_2 = 2\epsilon_1+\frac{\ddot{h}}{h\dot{h}} .
\end{equation}
The conditions for a slow-roll regime are satisfied when $\epsilon_1 < 1$ and $\epsilon_2 < 1$, 
and inflation ends when any of them exceeds unity.
The effect of the radion and the tachyon can be seen if we compare 
the slow-roll parameters for the full model with those for the model with inverse quartic tachyon potential.
The number of e-folds is defined through Eq.\ (\ref{n}) as 
\begin{equation}
 N=\int_{\tau_{\rm i}}^{\tau_{\rm f}}d\tau h,
\end{equation}
where the subscripts  ${\rm i}$ and ${\rm f}$ denote the beginning and the end of inflation, respectively.

The slow-roll parameters are related to observable quantities, in particular to 
the tensor-to-scalar ratio $r$ and the scalar spectral index $n_{\rm s}$
defined by
\begin{equation}
 r=\frac{\mathcal{P}_{\rm T}}{\mathcal{P}_{{\rm S}}},
 \label{eq3005}
 \end{equation}
\begin{equation}
 n_{\rm s}= \frac{d\ln \mathcal{P}_{{\rm S}}}{d\ln q},
 \label{eq3006}
 \end{equation}
 where $\mathcal{P}_{{\rm S}}$ and $\mathcal{P}_{\rm T}$ are the power spectra of scalar and tensor perturbations, respectively,
 evaluated at the horizon, i.e., for  a wave-number satisfying $q=aH$.
 Calculation of the spectra proceeds by identifying the proper canonical field and imposing 
  quantization of the quadratic action for the near free field. 
 The procedure for a general k-inflation is described in  \cite{garriga}
 and applied to the tachyon fluid in Refs. \cite{steer,frolov}.
 It turns out that at the lowest order in $\epsilon_1$ and $\epsilon_2$, the power spectra 
  may be expressed in the same way as in the standard tachyon inflation \cite{steer}:
 \begin{equation}
 \mathcal{P}_{\rm T} \simeq [1-2(1+C)\epsilon_1]\frac{16GH^2}{\pi},
 \label{eq3007}
 \end{equation}
\begin{equation}
 \mathcal{P}_{{\rm S}} \simeq [1-2(1+C-\alpha)\epsilon_1-C \epsilon_2]\frac{GH^2}{\pi\epsilon_1 },
\label{eq3008} 
\end{equation}
 where $C=-2+\ln 2 +\gamma\simeq -0.72$, and $\alpha$ is a parameter related to the speed of sound
 expanded in $\epsilon_1$:
 \begin{equation}
  c_{\rm s}=1-2\alpha\epsilon_1 +O(\epsilon_1^2).
  \label{eq3009}
 \end{equation}
The parameter $\alpha$ will be calculated in the next section using
the relation between $\epsilon_1$ and the speed of sound in the slow-roll approximation.

\section{Conditions for tachyon inflation}
\label{conditions}
\subsection{Inverse quartic potential}
To solve the system of equations (\ref{sysRT1})-(\ref{sysRT4}) 
we need to choose initial conditions relevant for inflation.
To this end we first solve a simpler case in which the radion is absent. 
The model is described by the pure tachyon Lagrangian with the inverse quartic potential \cite{bpu9}
\begin{equation}
{\cal{L}} =  
-\frac{\lambda}{\Theta^4}\sqrt{1-g^{\mu\nu}\Theta_{,\mu}\Theta_{,\nu}}  \,,
 \label{eq000}
\end{equation}
with the corresponding Hamiltonian  
\begin{equation}
{\cal{H}} =\frac{\lambda}{\Theta^4}\sqrt{1+\Pi_{\Theta}^2\Theta^8/(\lambda^2)}.
 \label{eq001}
\end{equation}
As before, we rescale the time as $t=\tau/k$,  
absorb the coupling constant $\lambda$ into the conjugate field $\Pi_{\Theta}$
 and introduce  the dimensionless functions as in (\ref{eq002}).
Then, using  (\ref{eqHam2}), (\ref{eqHam4}) and (\ref{eq001}) we obtain the Hamilton equations in the form
\begin{equation}
\dot \theta=\frac{\theta^4\pi_{\theta}}
{\sqrt{1+\theta^8\pi_{\theta}^2}}
\label{eq003}
\end{equation}
\begin{equation}
\dot \pi_\theta=-3h\pi_\theta
+\frac{4}{\theta^5
\sqrt{1+\theta^8\pi_\theta^2}}
\label{eq004}
\end{equation}
The dimensionless pressure and energy density are given by 
\begin{equation}
 \bar{p}= -\frac{1}{\theta^4\sqrt{1+\theta^8\pi_{\theta}^2}}=
 -\frac{1}{\theta^4}\sqrt{1-\dot{\theta}^2},
 \label{eq008}
\end{equation}
\begin{equation}
 \bar{\rho}= \frac{1}{\theta^4}\sqrt{1+\theta^8\pi_{\theta}^2}=
 \frac{1}{\theta^4}\frac{1}{\sqrt{1-\dot{\theta}^2}},
 \label{eq2008}
\end{equation}
Using this we also find a simple  expression for the sound speed
\begin{equation}
 c_{\rm s}^2\equiv \left.\frac{\partial p}{\partial\rho}\right|_{\theta} =\frac{1}{1+\theta^8\pi_{\theta}^2}=
 1-\dot{\theta}^2.
 \label{eq2009}
\end{equation}
The Friedmann equations are  given by Eq.\ (\ref{h}) and (\ref{h2}) with (\ref{eq008}) and (\ref{eq2008}).
Hence, as before,  the coupling $\lambda$ drops out and the only remaining parameter is
 $\kappa$ defined in (\ref{eq102}).
The system of equations (\ref{eq003})-(\ref{eq004}) can be solved numerically for a chosen set of initial conditions.
Once the solution for $H$ is found the slow-roll parameters can be easily calculated.

\subsection{Slow-roll approximation and initial conditions}
\label{initial}

To find appropriate initial value of the field $\theta$, we first consider the pure tachyon model in the slow-roll
approximation.
Tachyon inflation is based upon the slow evolution of $\theta$ with the slow-roll
conditions \cite{steer}
\begin{equation}
 \dot{\theta}\simeq \theta^4\pi_\theta \ll 1, \quad \dot\pi_\theta \ll 3h \pi_\theta,
\end{equation}
so that in the slow-roll approximation we may neglect the factors $(1-\dot{\theta}^2)^{1/2}$.
Then, during inflation we have
\begin{equation}
h\simeq \frac{\kappa}{\sqrt3 \theta^2}\left(1+\frac{\kappa^2}{12\theta^4}\right)^{1/2},
\label{eq1007}
\end{equation}
\begin{equation}
 \dot{\theta} \simeq \frac{4}{3h\theta} \simeq \frac{4\theta }{\sqrt3 \kappa}
\left(1+\frac{\kappa^2}{12\theta^4}\right)^{-1/2} ,
\label{eq1008}
\end{equation}
\begin{equation}
 \ddot{\theta} \simeq \frac{4\dot{\theta} }{\sqrt3 \kappa}
\left(1+\frac{\kappa^2}{12\theta^4}\right)^{-3/2} \left(1+\frac{\kappa^2}{4\theta^4}\right),
\label{eq1009}
\end{equation}
and using (\ref{h2}) we also find
\begin{equation}
 \dot{h}=-\frac{\kappa^2 \dot{\theta}^2}{2\theta^4}\left(1+\frac{\kappa^2}{6\theta^4}\right).
\end{equation}
As a consequence, the slow-roll parameters (\ref{eps1}) and (\ref{eps2}) can be approximated by  
\begin{eqnarray}
\epsilon_1&\simeq&\frac32 \dot{\theta}^2\left(1+\frac{\kappa^2}{6\theta^4}\right)
\left(1+\frac{\kappa^2}{12\theta^4}\right)^{-1}
\nonumber \\
&\simeq&\frac{8\theta^2}{\kappa^2}\left(1+\frac{\kappa^2}{6\theta^4}\right)
\left(1+\frac{\kappa^2}{12\theta^4}\right)^{-2},
\label{eq009}
\end{eqnarray}
\begin{eqnarray}
\epsilon_2 &\simeq& 2\frac{\ddot{\theta}}{h\dot{\theta}}
-\dot{\theta}^2\frac{\kappa^2}{4\theta^4}
\left(1+\frac{\kappa^2}{6\theta^4}\right)^{-1}\left(1+\frac{\kappa^2}{12\theta^4}\right)^{-1}
\nonumber \\ 
&\simeq&  \frac{8\theta^2}{\kappa^2} \left(1+\frac{\kappa^2}{12\theta^4}\right)^{-2}\left[
1+\frac{\kappa^2}{4\theta^4}-\frac{\kappa^2}{6\theta^4}\left(1+\frac{\kappa^2}{6\theta^4}\right)^{-1}
\right].
\label{eq010}
\end{eqnarray}
In the slow roll regime we have
$\kappa^2/\theta^4\gg 1$, so the corrections due to the RSII modification in Eqs.\ (\ref{eq1007})-(\ref{eq010}) 
will dominate over unity
and we find
\begin{equation}
h\simeq \frac16 \frac{\kappa^2}{\theta^4},
\quad\quad
\dot{\theta}\simeq 8 \frac{\theta^3 }{ \kappa^2} ,
\quad\quad
\ddot{\theta}\simeq 24 \frac{\theta^2}{\kappa^2}\dot{\theta},
 \label{eq021}
\end{equation}
\begin{equation}
\epsilon_1\simeq 3 \dot{\theta}^2\simeq 192 \frac{\theta^6}{\kappa^4},
\quad\quad
\epsilon_2\simeq 288 \frac{\theta^6}{\kappa^4} \simeq \frac32\epsilon_1 .
 \label{eq018}
\end{equation}
In contrast,
if we disregarded  the RSII corrections we would obtain the usual slow-roll equations
of tachyon inflation
\cite{steer,campo,li,garousi}
for the  potential $V=\lambda/ \theta^{4}$
\begin{equation}
h\simeq \frac{1}{\sqrt{3}} \frac{\kappa}{\theta^2},
\quad\quad
\dot{\theta}\simeq \frac{4}{\sqrt{3}} \frac{\theta}{ \kappa},
\quad\quad
\ddot{\theta}\simeq \frac{4}{\sqrt{3}} \frac{\dot{\theta}}{\kappa},
 \label{eq0021}
\end{equation}
\begin{equation}
\epsilon_1\simeq \frac32 \dot{\theta}^2\simeq 8 \frac{\theta^2}{\kappa^2},
\quad\quad
\epsilon_2\simeq \epsilon_1.
 \label{eq0018}
\end{equation}
Hence, in the slow-roll regime the tachyon inflation in the RSII modified cosmology proceeds in a quite distinct
way compared with that in the standard FRW cosmology. However,
close to and at the end of inflation we have $\kappa^2/\theta_{\rm f}^4\ll 1$ and we 
can neglect the RSII cosmology corrections. 
Hence, the expressions (\ref{eq0021}) and (\ref{eq0018})
can be used at the end of inflation where 
we find 
\begin{equation}
 \epsilon_1(\theta_{\rm f})\simeq \epsilon_2(\theta_{\rm f})\simeq \frac{8\theta_{\rm f}^2}{\kappa^2}
 \simeq 1,
\label{eq017} 
 \end{equation}
 and 
\begin{equation}
h(\theta_{\rm f})\simeq \frac{8}{\sqrt3 \kappa}.
\label{eq2007}
\end{equation} 
 
In the slow-roll approximation  the number of e-folds  is given by 
\begin{eqnarray}\label{inteq}
N \simeq\frac{\kappa^2}{4} \int_{\theta_0}^{\theta_{\rm{f}}} \frac{d\theta}{\theta^3}
\left(1+\frac{\kappa^2}{12\theta^4}\right) \simeq
\frac{\kappa^2}{8\theta_0^2}\left(1+\frac{\kappa^2}{36\theta_0^4}\right)-1 
\simeq \frac23 \frac{1}{\epsilon_1(\theta_0)}-1,
\end{eqnarray}
where we  have exploited $\kappa^2/(36\theta_0^4)\gg 1 $ at the beginning 
 and $\kappa^2/(36\theta_{\rm f}^4)\ll 1 $ at the end of inflation together with
the condition  (\ref{eq017}).
For comparison, in the standard tachyon inflation 
described by (\ref{eq0021}) and (\ref{eq0018}) 
we would obtain 
\begin{equation}
N_{\rm st.tach} \simeq\frac{\kappa^2}{8\theta_0^2}-1 \simeq \frac{1}{\epsilon_1(\theta_0)}-1.
 \label{eq319}
\end{equation}
For example, the choice $\kappa^2=5$ and  $\theta_0 = 0.25$  
leads to $N_{\rm st.tach} \simeq 9 $ whereas in RSII cosmology with the same parameters
one finds $N \simeq 365$.

At this point it is convenient to estimate a phenomenologically acceptable range of the couplings $\lambda$
and $\kappa$.
Although the evolution equations do not depend on $\lambda$,
its approximate value is needed for
choosing  appropriate initial conditions for the radion field.
The value of $\lambda$ 
 may be estimated using the 
observational constraint on the amplitude of scalar perturbations.
The approximate expression
\begin{equation}
 \mathcal{P}_{\rm S}\simeq \frac{GH^2}{\pi\epsilon_1},
\end{equation}
which follows from (\ref{eq3008}),
 is to be compared  with the power spectrum amplitude $A_s$ measured by Planck \cite{planck2015}:
\begin{equation}
 A_s\simeq 2.2\times 10^{-9}.
\end{equation}
Hence,  we must make sure that the condition 
\begin{equation}
 \frac{H}{M_{\rm P}} \lesssim \sqrt{\pi A_s}\simeq 8.31\times 10^{-5}
\end{equation}
is satisfied close to and at the end of inflation (where $\epsilon_1 \lesssim 1$). 
(See also Ref.\ \cite{thomas}.)
Here we define the Planck mass as  $M_{\rm P}=G^{-1/2}$.
According to Eq.\ (\ref{eq2007}) in  the slow-roll approximation near the end of inflation we have
\begin{equation}
 \frac{H}{M_{\rm P}}\simeq \frac{8}{\sqrt{3}} \frac{k}{\kappa M_{\rm P}}=
\sqrt{\frac{8 }{3\pi}}\frac{k^2}{\sqrt{\sigma}} ,
\label{eq03}
\end{equation}
yielding
\begin{equation}
 \frac{k}{\sigma^{1/4}}\lesssim  10^{-2},
 \label{eq01}
\end{equation}
or
\begin{equation}
 \lambda\gtrsim  10^{8}.
 \label{eq02}
\end{equation}

To estimate $\kappa$ note first that the tension of a D$p$-brane is given by \cite{johnson}
\begin{equation}
 \sigma =\frac{1}{(2\pi)^p \alpha^{\prime (p+1)/2} g_{\rm s}},
\end{equation}
where $g_{\rm s}$ is the string coupling constant and $1/(2\pi\alpha^{\prime})$ is the string tension.
Using this for $p=3$ from (\ref{eq01}) we find a constraint 
\begin{equation}
 g_{\rm s}\lesssim 4\times 10^{-11}\left(\frac{M_{\rm s}}{k}\right)^4 ,
\end{equation}
where $M_{\rm s}=1/\sqrt{\alpha^{\prime}}$. 
Using this constraint we can choose $k$ and $M_{\rm s}$ such that the 
scale hierarchy 
\begin{equation}
 H < k < M_{\rm s} < M_{\rm P}
 \end{equation}
is satisfied.
With this in mind we can give an order of magnitude estimate 
for an acceptable range of values of our free parameter $\kappa$.
Requiring  $k\gtrsim H$, from (\ref{eq03}) it follows 
\begin{equation}
 \kappa\gtrsim 8/\sqrt{3},
 \end{equation}
so we can safely choose $\kappa$ to vary in the range $1<\kappa<20$.

Equations (\ref{eq018}) and (\ref{inteq}) can be used to estimate the value of the ratio $\kappa^2/\theta_0^3$ by fitting the model 
to the observational parameters $r$ and $n_{\rm s}$ defined by (\ref{eq3005}) and (\ref{eq3006}). 
First, using (\ref{eq2009}) and  (\ref{eq018})  we can express the sound speed in terms of $\epsilon_1$: 
\begin{equation}
 c_s \simeq1-\frac16 \epsilon_1,
\end{equation}
and comparing this with (\ref{eq3009})
we find 
$\alpha=1/12$.
In contrast, in the standard tachyon inflation
 the value  $\alpha=1/6$ is obtained \cite{steer}. 
 Using (\ref{eq3007}) and (\ref{eq3008})
we find  up to the second order 
 \begin{equation}
r=16 \epsilon_1\left[1- \frac16 \epsilon_1+C\epsilon_2 \right],
\label{eq005}
\end{equation}
\begin{equation}
n_{\rm s}=1-2 \epsilon_1 - \epsilon_2
-\left[2 \epsilon_1^2+\left(2C +\frac{17}{6} \right)\epsilon_1\epsilon_2 
+C\epsilon_2\epsilon_3\right],
\label{eq006}
\end{equation}
where it is understood that  $\epsilon_1$ and $\epsilon_2$  take their values at or close to the beginning of inflation.
Using these equations        
at linear order with $\epsilon_2\simeq 3\epsilon_1/2$
and (\ref{inteq}),
we find the approximate relations
\begin{equation}
 r=\frac{32}{3}\frac{1}{N+1} ,
 \label{eq501}
\end{equation}
\begin{equation}
 n_{\rm s}=1-\frac73 \frac{1}{N+1},
  \label{eq502}
\end{equation}
\begin{equation}
 r=\frac{32}{7}(1-n_{\rm s}).
  \label{eq503}
 \end{equation}
Had we used the usual relation $\epsilon_2\simeq \epsilon_1$
and (\ref{eq319}) we would have obtained the standard tachyon-inflation relations \cite{steer,li}
\begin{equation}
 r=\frac{16}{N+1},
\end{equation}
\begin{equation}
 n_{\rm s}=1-\frac{3}{N+1},
\end{equation}
\begin{equation}
 r=\frac{16}{3} (1-n_{\rm s}).
 \label{eq5004}
\end{equation}
 
Comparing $r$ and $n_{\rm s}$  with the latest  observations we can fix the  parameters  
 $\epsilon_1$, $\epsilon_2$, and $N$. Then,
from (\ref{inteq}) we can  determine the ratio $\kappa^2/\theta_0^3$
and for a chosen set of values of $\kappa$
we  find the corresponding  $\theta_0$.
These values of $\kappa$ and the corresponding initial values $\theta_0$ as initial conditions
are then used to solve the system of equation (\ref{eq003})-(\ref{eq004}) for the
pure tachyon and (\ref{sysRT1})-(\ref{sysRT4}) for the tachyon-radion system.

There is no a priori reason to restrict possible  initial values of the radion field $\Phi$ 
so we can choose a range of initial values 
based on the natural scale dictated by observations, i.e., the scale 
between $H$ and $M_{\rm P}$. Hence, a natural initial value for $\Phi$ would 
be of the order of $k$ or a few orders larger, say in the range
10 to 1000 $k$.  
However, according to (\ref{eq002}) the dimensionless radion field $\phi$ is rescaled with respect to $\Phi/k$
by a factor of $1/\sqrt{\lambda}\simeq 10^{-4}$ so
we can choose the initial value $\phi_0=\phi(0)$
in the range from 0.001 to 0.5.

As we have mentioned in Introduction our tachyon model suffers from the so called reheating problem
\cite{kofman}. A possible way out is provided by string theory.
String theory D-branes couple to the (pull-back of)
antisymmetric tensor field $\mathcal{B}$ that combines the Kalb-Ramond and electromagnetic fields.
In this way there exist a natural coupling between the tachyon and the electromagnetic field.
Therefore,  this interaction  could serve as a possible reheating mechanism at the end of inflation.
In a tachyon model based on brane-antibrane annihilation
resulting in a time dependent tachyon condensate
it has been shown \cite{cline} that a coupling of massless fields to the time dependent 
tachyon condensate could yield a reheating efficient enough to overcome the above mention problem
of a CDM dominance. 

Another possible way out of the reheating problem could be the so called {\em warm inflation}
\cite{berera}. Warm inflation is an alternative inflation scenario with
no need for a reheating period. In warm inflation, dissipative effects are included
during inflation, so that radiation is produced in parallel with the inflationary
expansion and  inflation ends when the universe heats up to become radiation dominated.
This scenario has been successfully applied to tachyon inflation models
\cite{herrera,motaharfar} and, in principle, should also work for our model.
This requires further investigation which goes beyond the scope of the present paper.

\section{Numerical results}\label{Sec:numres}
\begin{figure}[!ht]

\includegraphics[width=0.45\textwidth,trim= 0 0cm 0 0cm]{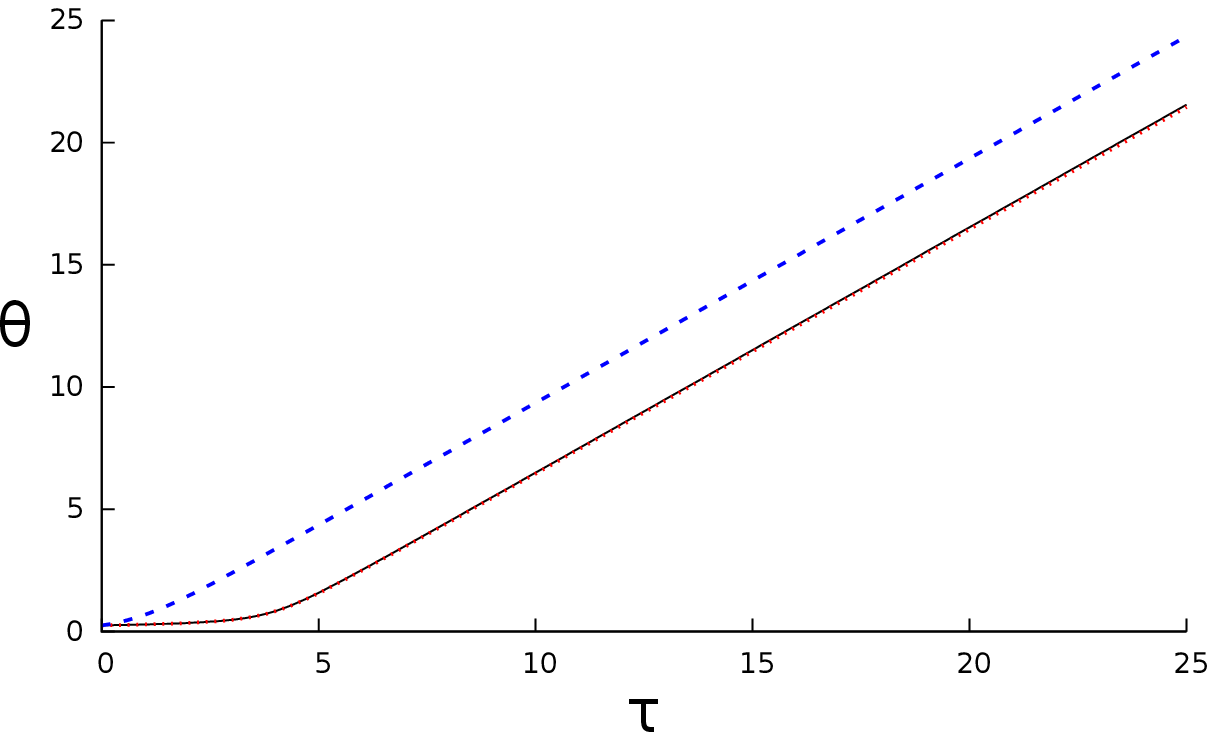}
\hspace{0.02\textwidth}
\includegraphics[width=0.45\textwidth]{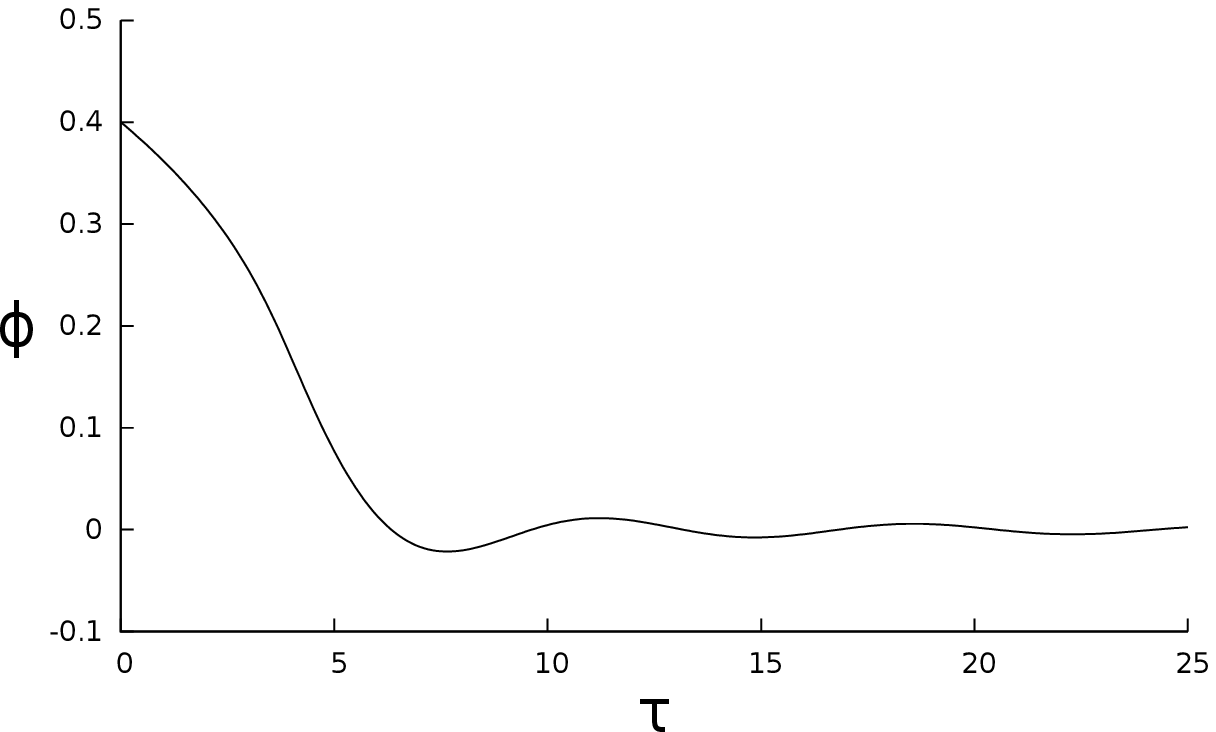}
\\
\vspace{0.02\textwidth}
\includegraphics[width=0.45\textwidth]{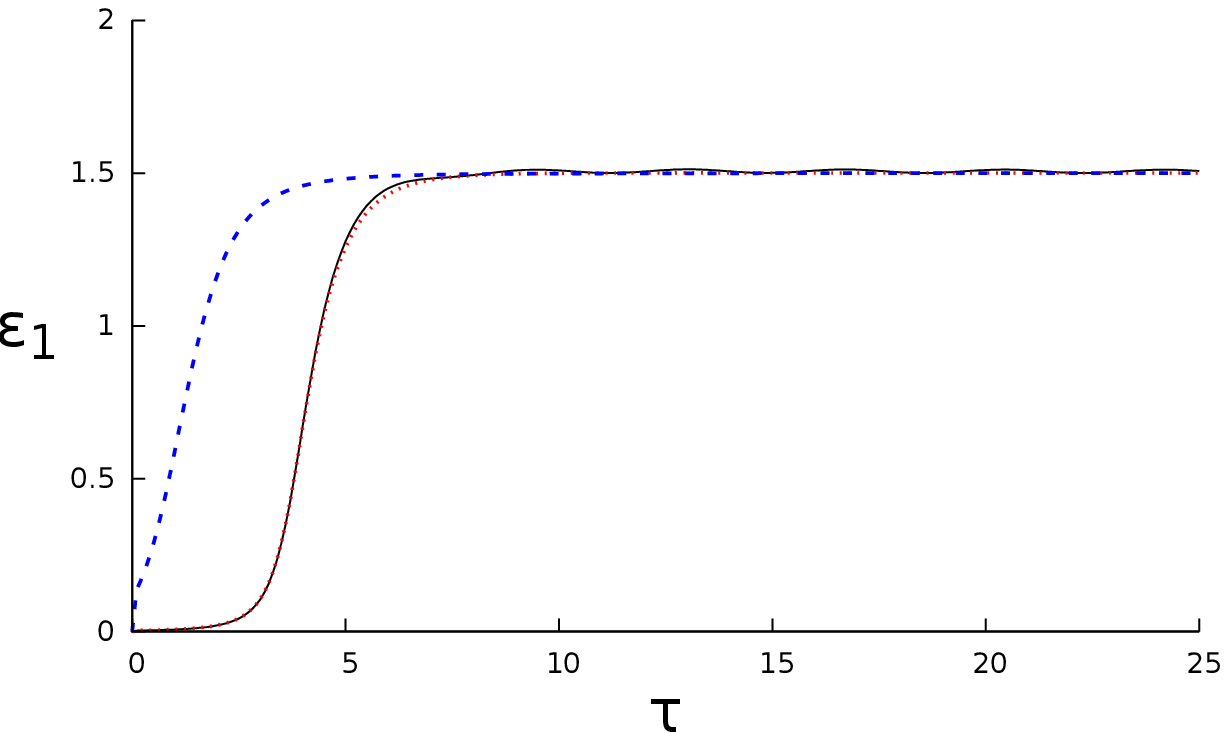}
\hspace{0.02\textwidth}
\includegraphics[width=0.45\textwidth]{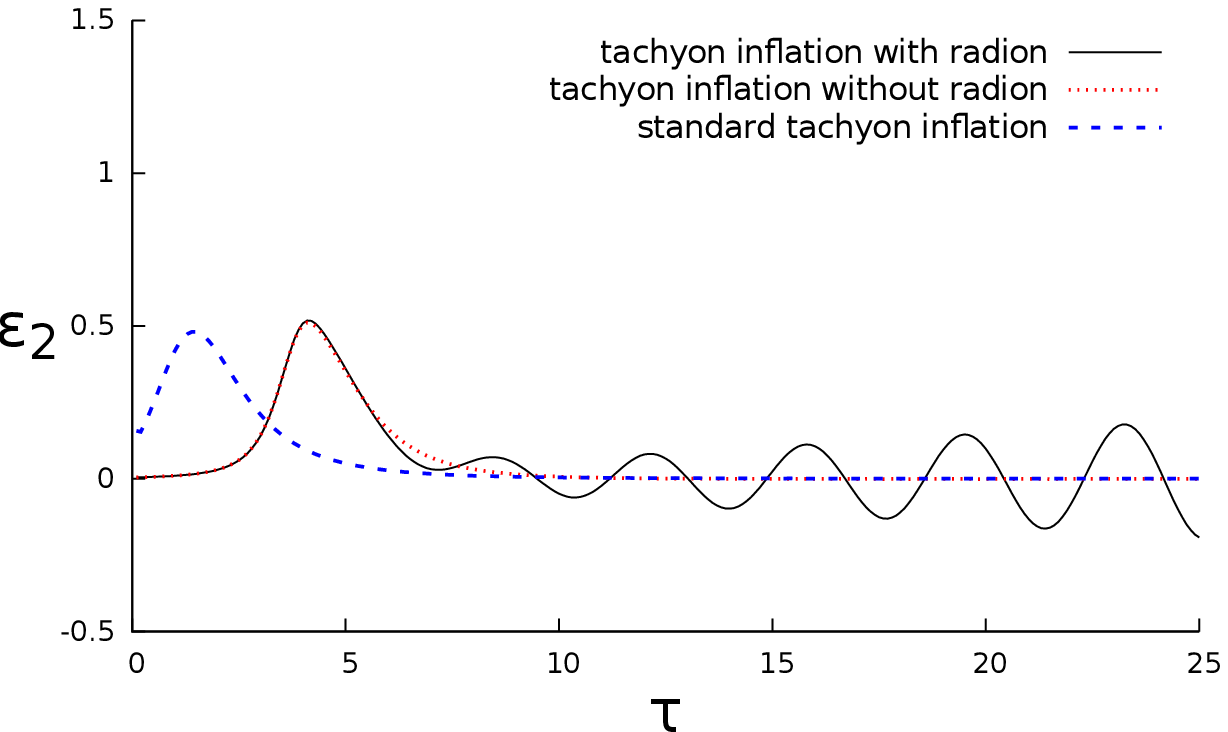}
\caption{Time evolution of the tachyon $\theta$ (top left), 
the radion $\phi$ (top right) fields  and  of the slow-roll parameters $\epsilon_1$ (bottom left) and 
$\epsilon_2$ (bottom right) for 
  $\kappa=2$, $\phi_0=0.4$, $\pi_{\theta 0}=\pi_{\phi 0}=0$, $\theta_0=0.25$, in the tachyon-inflation
model with inverse quartic potential in the RSII cosmology  with radion (full black line) and without radion (dotted red line).
The dashed blue line represents the corresponding results for the  tachyon-inflation model with inverse quartic potential 
in the standard FRW cosmology
and no radion.}
 \label{fig1}
\end{figure}

The system of equations (\ref{sysRT1})-(\ref{sysRT4}) is evolved numerically starting from $\tau=0$
 up to some large $\tau$ of the order of 100.
 The initial values $\theta_0$ and $\phi_0$ are chosen as described in the previous section and
the  initial conjugate momenta are taken 
to be $\pi_{\theta 0}=\pi_{\phi 0}=0$. 
The function $N(\tau)$  is solved simultaneously using
(\ref{n}) with $N(0)=0$.
The time evolution of the slow roll parameters $\epsilon_1$ and $\epsilon_2$ are obtained  
using (\ref{eps1}) and (\ref{eps2}).
The  inflation ends at a point $\tau_{\rm f}$ at which $\epsilon_1(\tau_{\rm f})=1$.
The  beginning of inflation at $\tau_i$ is then found by requiring 
$N(\tau_{\rm f})-N(\tau_{\rm i})=N$.
The results for $\kappa=2$, $\theta_0=0.25$ and $\phi_0=0.4$ are presented in Fig.\ \ref{fig1}
 together with the results calculated for the system (\ref{eq003})-(\ref{eq004})
 within the standard FRW cosmology
with the same initial $\theta_0=0.25$ and $\pi_{\theta 0}=0$.

\begin{figure}[ht!]
\begin{center}
\includegraphics[width=0.45\textwidth,trim= 0 0cm 0 0cm]{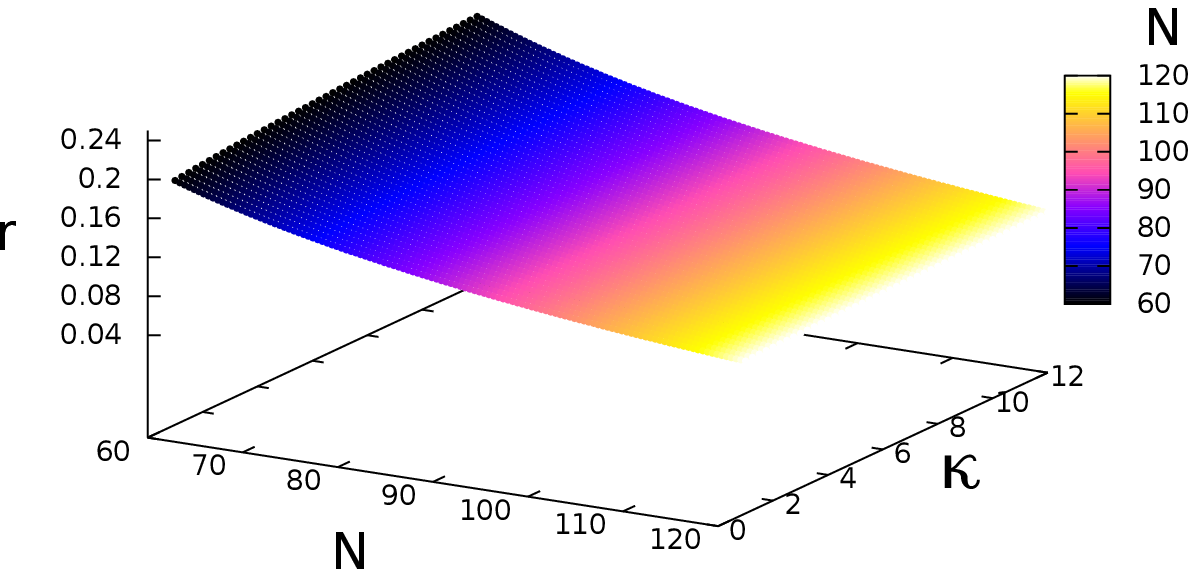}
\hspace{0.02\textwidth}
\includegraphics[width=0.45\textwidth]{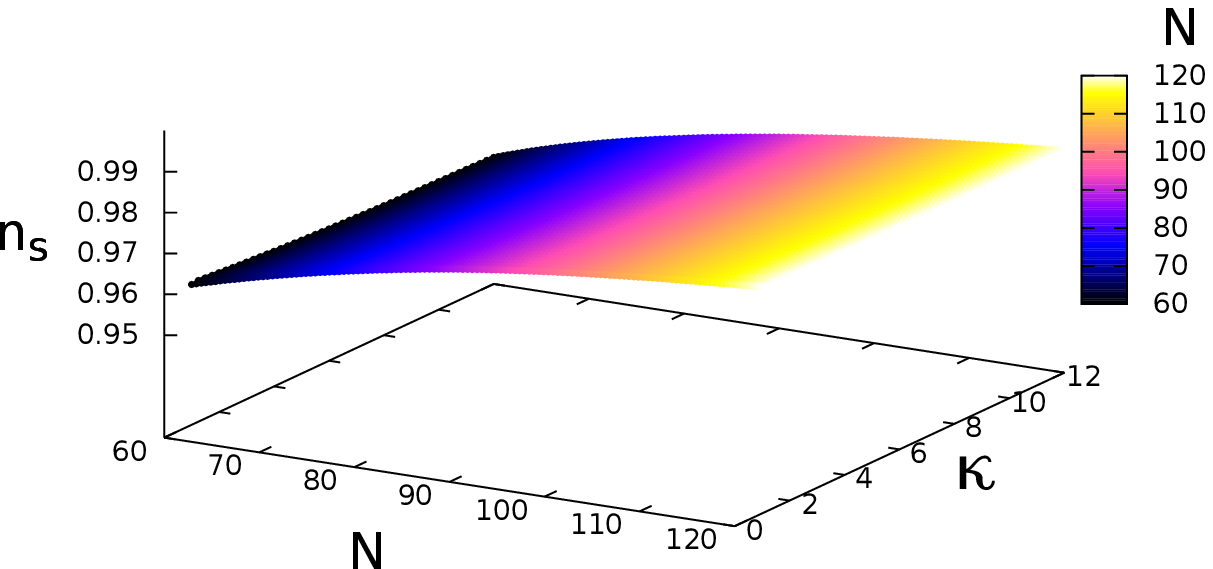}
\caption{The quantities $r$ (left) and $n_{\rm s}$ (right) as functions 
of the parameters $N$ and $\kappa$ as indicated on the horizontal axes for initial $\phi_0=0.01$. 
The initial $\theta_0$ varies with $N$ and $\kappa$ according to Eq.\ (\ref{inteq}).}
 \label{fig2}
\end{center}
\end{figure}

To calculate the quantities $n_s$ and $r$ we proceed as follows. 
For a chosen pair of $(N, \kappa)$ we first find the initial value $\theta_0$ from Eq.\ (\ref{inteq}). 
Then, for a  chosen set of initial values $\phi_0$, $\pi_{\theta 0}$, $\pi_{\phi 0}$ (at $\tau=0$) 
we find the corresponding $\phi_{\rm i}$, $\pi_{\theta \rm i}$, $\pi_{\phi \rm i}$ at $\tau=\tau_{\rm i}$ 
and calculate $h$, $\dot{h}$ and $\ddot{h}$ using equations (\ref{sysRT1})-(\ref{h}) with (\ref{eq504}) 
and (\ref{eq505}). 
From this we find $\epsilon_1(\theta_{\rm{i}})$ and $\epsilon_2(\theta_{\rm{i}})$ 
using the defining expressions (\ref{eps1}) and (\ref{eps2}) and 
calculate the parameters $r$ and $n_{\rm s}$ from (\ref{eq005}) and (\ref{eq006}), respectively. 
In Fig.\ \ref{fig2} we present $n_s$ and $r$ as functions of $\kappa$ and $N$ for a fixed $\phi_0=0.01$. 

\begin{figure}[ht!]
  \centering
\includegraphics[width=0.5\textwidth,trim= 0 0cm 0 0cm]{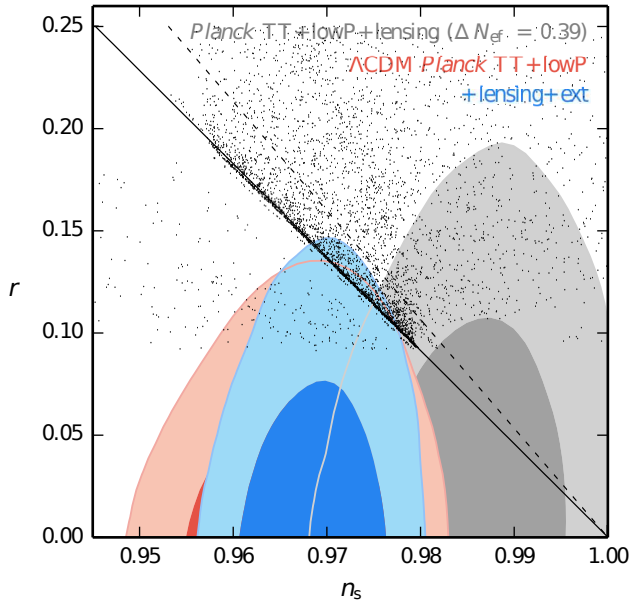}
  \caption{$r$ versus $n_{\rm s}$ diagram with observational constraints 
from Ref.\  \cite{planck2015}.
The dots represent the calculation
in the tachyon-radion model 
for various $N$, $\kappa$ and $\phi_0$ chosen randomly in the range $60 \leq N \leq 120$, $1\leq \kappa \leq 12$ and $0\leq \phi_0 \leq 0.5$. The full line represents 
 the slow-roll approximation (Eq.\ (\ref{eq503})) of the RSII model with no radion.
The dashed line represents the slow-roll approximation (Eq.\ (\ref{eq5004})) of 
the standard tachyon model with inverse quartic potential.} 
\label{fig3}
\end{figure}

In Fig.\ \ref{fig3} we present the $n_s-r$ diagram with 10000 points superimposed 
on the observational constraints taken from the Planck Collaboration 2015 \cite{planck2015}. 
Each point in the diagram corresponds to a set $(N, \kappa, \phi_0)$ chosen randomly 
in the range $60 \leq N \leq 120$, $1\leq \kappa \leq 12$ and $0\leq \phi_0 \leq 0.5$. 

\begin{figure}[ht!]
  \centering
\includegraphics[width=0.5\textwidth,height=0.5\textwidth]{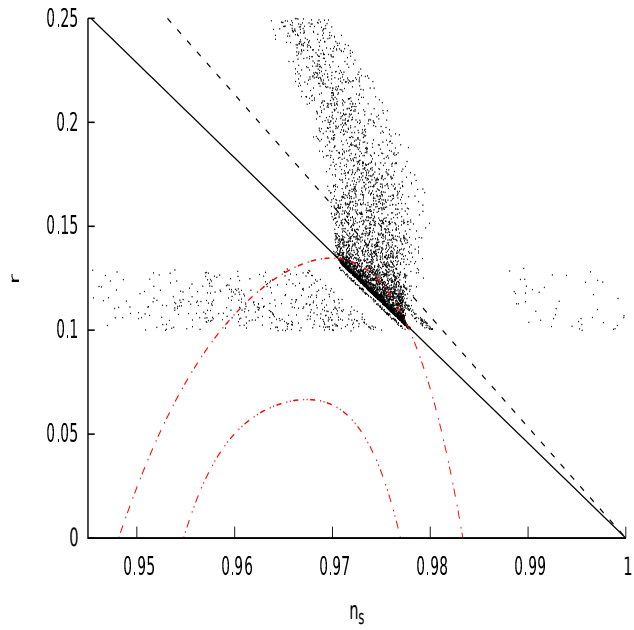}
  \caption{Same as in Fig.\ \ref{fig3}  for  $N$, $\kappa$, and $\phi_0$ 
chosen randomly in the intervals $85 \leq N \leq 110$, $1\leq \kappa \leq 8$ and $0\leq \phi_0 \leq 0.5$. 
The inner and outer dash-dotted lines denote the Planck  contours of the one and two $\sigma$ constraints, respectively.
} 
\label{fig4}
\end{figure}

To obtain a more favorable distribution of points  within the 2$\sigma$ area measured by the Planck collaboration
we have used the distribution histograms
of the number of e-folds $N$ and the parameters $\kappa$ and $\phi_0$.
In this way we have been guided  to restrict 
  $N$  to range between 85 and 110, 
 $\kappa$ between 1 and 8,  
and  $\phi_0$ between 0 and 0.5. The outcome of these constraints is presented in Fig.\ \ref{fig4}. 
In this figure one notices basically two bands: 
the dominant one almost parallel to the $r$ axis with most data points concentrated close to 
the line corresponding to the RSII model  prediction without radion 
 and the less dens one almost parallel to the $n_s$ axis, with $r$ between 0.1 and 0.13.

\begin{figure}[ht!]
  \centering
\includegraphics[width=0.5\textwidth,height=0.5\textwidth]{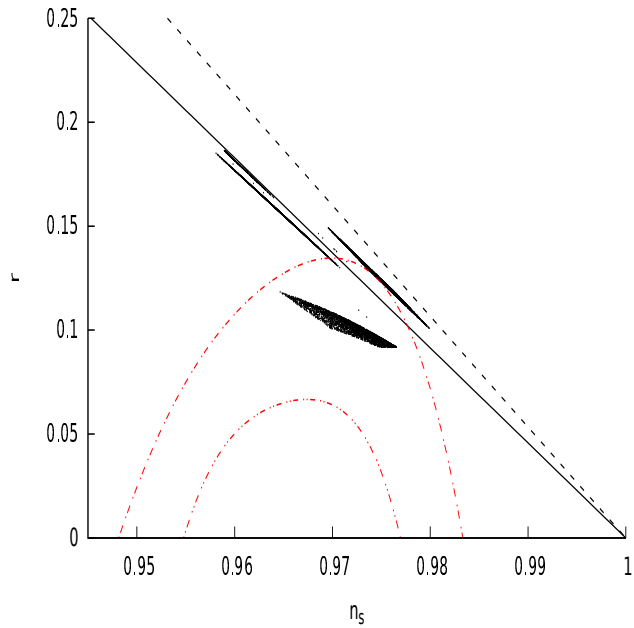}
  \caption{Same as in Fig.\ \ref{fig4}  for a fixed $\kappa=2$ and $N$ and $\phi_0$
  chosen randomly in the intervals $60 \leq N \leq 120$ and $0\leq \phi_0 \leq 0.25$.} 
\label{fig5}
\end{figure}

From a quite large data set of numerical results we find that smaller values of $\kappa$ 
give  a better agreement with observational data from the Planck mission. 
For example, if we fix $\kappa=2$ 
and vary $N$ in a wider range  $N$  $60\leq N \leq 120$) and $\phi_0$ in  a bit narrower range $0 \leq\phi_0 \leq 0.25$)
we obtain an interesting splitting of points into three disconnected clusters (Fig.\ \ref{fig5}).
 This splitting is a clear manifestation of the nonlinearity of Eqs.\ (\ref{sysRT1})-(\ref{sysRT4}). 
The most pronounced cluster of numerical 
points fits in quite well within the 2$\sigma$ regions of constraints given by the Planck collaboration.

\begin{figure}[ht!]
  \centering
\includegraphics[width=0.5\textwidth,height=0.5\textwidth]{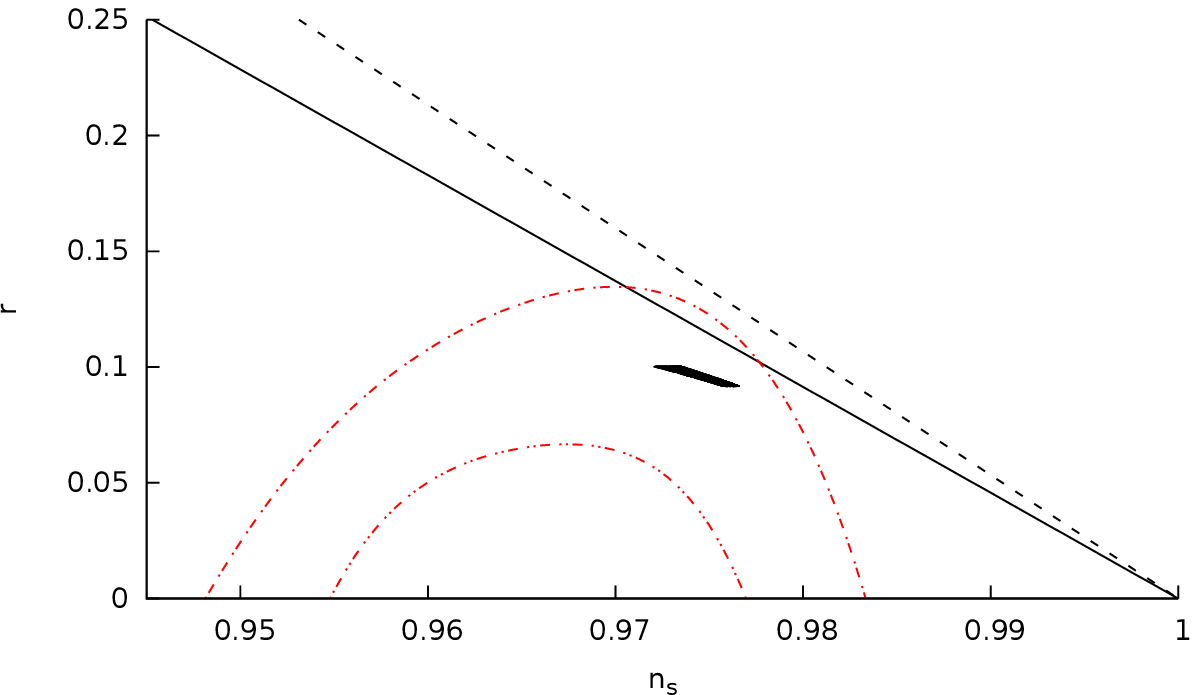}
  \caption{Same as in Fig.\ \ref{fig5} for  $N$ 
  and $\phi_0$ chosen randomly in the intervals $110 \leq N \leq 120$ and $0.1\leq \phi_0 \leq 0.2$.} 
\label{fig6}
\end{figure}

\begin{figure}[ht!]
  \centering
\includegraphics[width=0.5\textwidth,height=0.5\textwidth]{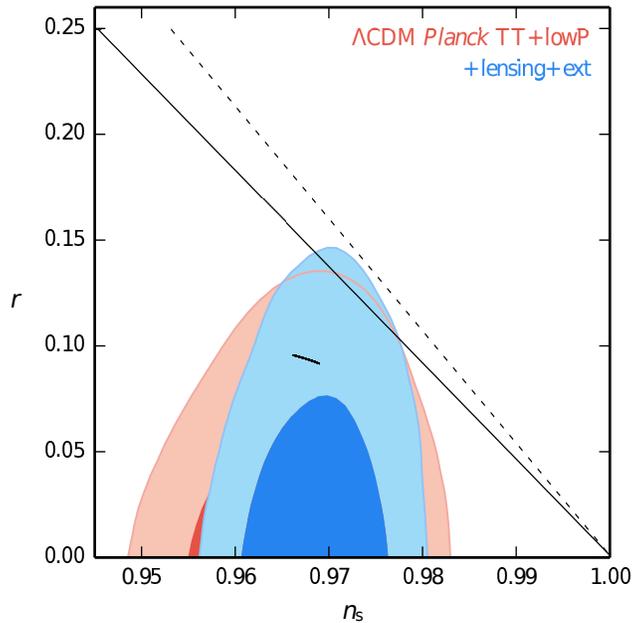}
  \caption{Same as in Fig.\ \ref{fig3} for fixed $\kappa=1.25$ and  $\phi_0=0.05$ and $N$ 
  chosen randomly in the interval $115 \leq N \leq 120$.} 
\label{fig7}
\end{figure}

A still  better agreement between our model and the observational data constraints
is obtained for  a very narrow range of parameters as shown in Figs.\ \ref{fig6} and \ref{fig7}.
The  best fit is obtained for $\kappa=1.25$, $\phi_0=0.05$ and $115 \leq N \leq 120$.

\section{Summary and conclusions}
\label{conclusions}
We have investigated a model of inflation based on the dynamics of a D3-brane 
in the AdS$_5$ bulk of the RSII model. The bulk metric is extended to include the back reaction of the radion
excitations.

We have shown that the slow-roll equations of the tachyon inflation are quite distinct to those
of the standard tachyon inflation with the same potential. In particular, 
 the departure  of the  sound speed
from unity equals $c_{\rm s}-1\simeq -\epsilon_1/6$ in contrast to the standard result 
 $c_{\rm s}-1\simeq -\epsilon_1/3$.
 The $n_{\rm s}-r$ relation in our model is substantially different from the standard one and
 is closer to the best observational value, as shown in Fig.\ \ref{fig6}.
 Note that the largest concentration of numerical results, represented by the points in the plot
 in Fig.\ \ref{fig6}, are within 2$\sigma$ of the {\em Planck} TT+lowP and {\em Planck}
TT+lowP+lensing\-+BAO+JLA+H$_0$ (red and light blue shaded regions, respectively).
Clearly, the agreement with observations is not ideal and it is fair to say that
the present  model is disfavored but not excluded.
 However, one should bare in mind that the model is based on the brane dynamics which results in a definite potential
 with one free parameter only. In contrast, the majority of other tachyon potentials discussed in
 the literature are chosen arbitrarily and adjusted so that the results agree with observations.
 
In this work we have analyzed the simplest tachyon model that stems from
the dynamics of a D3-brane in an AdS$_5$ bulk yielding basically an inverse quartic
potential.
In principle, the same mechanism could lead to a more general tachyon potential
if the AdS$_5$ background metric is deformed by 
the presence of matter in the bulk, e.g.,
in the form of a minimally coupled scalar field with an arbitrary 
self-interaction potential.
This will be the subject of our investigation in the future.
 
\section*{Acknowledgments}
This work has been supported by ICTP - SEENET-MTP project PRJ-09 Cosmology and Strings.
The work of N.\ Bili\'c has been supported by the Croatian 
Science Foundation under the project (IP-2014-09-9582).
D.\ Dimitrijevic, G.\ Djordjevic and M.\ Milosevic 
acknowledge support provided by the Serbian Ministry for Education, 
Science and Technological Development under the projects No 176021 
and No 174020. G.\ Djordjevic would like to thank CERN-Theoretical Physics Department 
for kind hospitality during the finalization of this paper.


\end{document}